\documentclass[proceedings, preprint]{rmaa}

\usepackage{paralist}
\usepackage{amsmath}
\usepackage{upgreek}

\usepackage{psfrag,color}

\SetYear{2022}
\SetConfTitle{Prospects for low-frequency radio astronomy in S. A.}

\title{An Argentinian window to the fast transient sky and to the very high resolution observations}

\author{
  B.~Marcote\altaffilmark{1}}

\altaffiltext{1}{Joint Institute for VLBI ERIC, Oude Hoogeveensedijk 4, 7991~PD Dwingeloo, The Netherlands (marcote@jive.eu).}

\shortauthor{Marcote}
\shorttitle{An Argentinian window to the transient sky}

\listofauthors{B. Marcote}
\indexauthor{Marcote, B.}

\abstract{The transient sky is composed of diverse phenomena that exhibits dramatic changes on short timescales. These events range from sub-second bursts to weeks and month timescale variability from compact systems. Several challenges need to be addressed by any facility that aims to observe such events: a fast re-positioning scheme to trace the first moments of events like Gamma-Ray Bursts (GRBs), a large field of view to be able to detect new Fast Radio Bursts (FRBs), or high sensitivity and high cadence to detect the outflows and flaring activity in Galactic binaries. Combined with a large bandwidth in order to recover the spectral information from these sources, it would allow us to unveil the physical processes taking place in these systems.

The new Multipurpose Interferometer Array (MIA) in Argentina may represent a suitable facility to conduct deep and leading-edge studies on the transient sky as the aforementioned ones. Additionally, there is a significant interest from the community on the possibility of connecting the 30-m IAR antennas within a VLBI network such as the European VLBI Network (EVN). This would place Argentina in the map to achieve very-high-resolution (on the milliarcsecond level) observations. This mode, together with the observations with the MIA would open a potential new regime that would allow astronomers to significantly increase the knowledge on the Southern Sky.}

\resumen{Existe un gran número de eventos en el Universo que presentan grandes cambios en escalas de tiempo muy cortas. Entre ellos, se pueden considerar los estallidos rápidos de radio (FRBs), que únicamente duran milisegundos; los {\it afterglow} de estallidos de rayos gamma (GRBs), que pueden durar meses; o la emisión de binarias de alta energía. Cualquier observatorio que quiera contribuir a observar este tipo de eventos deberá mostrar diversas cualidades: una velocidad de reposicionamiento suficientemente rápida para poder captar los primeros instantes de GRBs, un gran campo de visión para poder detectar un número reseñable de FRBs, y suficiente sensibilidad y ancho de banda como para restringir la emisión radio de diversos sistemas binarios.

El nuevo Multipurpose Interferometer Array (MIA) que se construirá en Argentina representa una instalación que podrá liderar la investigación de objetos transitorios en el hemisferio sur. De forma adicional, las actuales antenas de 30~m del IAR podrían contribuir a las observaciones de muy alta resolución (VLBI) como parte del European VLBI Network (EVN). La combinación de estos dos {\it arrays} abriría una nueva e importante ventana en el hemisferio sur para los astrónomos.}

\addkeyword{radiation mechanisms: non-thermal}
\addkeyword{techniques: high angular resolution}
\addkeyword{techniques: interferometers}
\addkeyword{radio continuum: general}

\begin{document}
\maketitle

\section{The Transient Sky}

The Universe is a highly dynamic and ever-evolving entity, constantly shaped by a multitude of powerful and cataclysmic events. From supernovae to the collision of binary neutron star systems, the Universe is a theater of constant change and transformation. It is within this dynamic framework that the study of the transient sky becomes crucial. By exploring the transient phenomena that is visible from our sky, we gain invaluable insights into the underlying extreme physical processes that take place in these astrophysical environments. Understanding the transient sky is therefore essential for unraveling the mysteries of the Universe and piecing together the intricate puzzle of cosmic evolution.

Most of these energetic phenomena involve particle acceleration up to relativistic energies, radiating non-thermal emission from radio to very-high-energy gamma rays.  Ground-based radio interferometers play a significant role in the exploration and understanding of these events as they show unparalleled angular resolution and sensitivity, that other facilities at higher energies cannot even relate. The Nature also collaborates with us in this point, as for most of these sources, the gigahertz frequency range seems to be a sweet spot where the non-thermal radio emission typically peaks. Roughly speaking, most of these transient sources would display a radio emission dominated by synchrotron processes, which result in an inverted power-law spectrum. However, at low frequencies ($\lesssim 1~\text{GHz}$), different absorption processes start to play a major role to suppress the observed emission \citep{longair2011}.
Therefore, the gigahertz frequency range allow us to observe these sources at the maximum sensitivity.

While the Northern hemisphere has always been well covered with different facilities observing at radio frequencies, the Southern hemisphere only presents a handful of instruments that can perform observations at gigahertz frequencies with enough sensitivity and resolution for our purposes.
The addition of new facilities is thus key to open the $2\pi$ window. Each of these facilities can thus boost our knowledge of the events being displayed at these negative latitudes.
The new observatories, operating or to be built, in South America would thus focused the interest of the community. The new Multipurpose Interferometer Array (MIA) would be a competitive instrument unique in the Southern hemisphere that could produce breakthrough discoveries in these fields. Additionally, the current 30-m antennas at the Argentine Institute of Radio astronomy (IAR, La Plata) can offer complementary observations and expand the range of the types of radio observations conducted in the continent.

In these proceedings I will just make a special focus on a few of the transient fields that are pushing our knowledge on the most extreme astrophysical regimes forward in the last decades. These fields are likely to provide a significant impact in the coming future for different areas, from fundamental physics to cosmology. This will serve as motivation for the possible use of the Argentinian instruments by the transient community.

\section{Galactic Gamma-Ray Emitting Binaries}

Several binary systems in our Galaxy have been found to display non-thermal emission that can reach up to high energy (HE; 0.1--100~GeV) and/or very high energy (VHE; $\gtrsim 100~\mathrm{GeV}$) gamma-rays \citep{dubus2015}. Not all these binaries belong to the same type of systems, and the mechanisms producing the gamma-ray emission are diverse among them. However, all these binaries can be catalogued into four groups: {\em novae} (the thermonuclear explosion on the surface of a white dwarf), {\em microquasars} (a neutron star or black hole accreting material from the donor star), {\em gamma-ray binaries} (relativistic wind collision between a putative young neutron star and the companion star winds), and {\em colliding wind binaries} (from the collision between the stellar winds of two massive stars).

In these proceedings I would just focus on the last two categories.

\subsection{Colliding Wind Binaries}

Massive stars (of O, B, or Wolf-Rayet spectral type) exhibit strong winds, with stellar wind velocities on the range of $\gtrsim 1\,000~\mathrm{km\ s^{-1}}$ and mass-loss rates of $\sim 10^{-4}\text{--}10^{-8}~\mathrm{M_\sun\ yr^{-1}}$ .
Interestingly, a large fraction of these massive stars are known to be part of binary or higher multiplicity systems. In such a configuration, the strong stellar winds collide, opening up valuable prospects for shock physics. In particular, the Diffusive Shock Acceleration (DSA) mechanism appears to be operative in tens of massive stellar systems, upstream of non-thermal emission processes. 
This subset of the massive star population is referred to as Particle-Accelerating Colliding-Wind Binaries \citep[PACWBs; ][]{debecker2013}.

A key property about this class of objects is the strong influence of free-free absorption (FFA) that is likely to prevent any detection of synchrotron radiation at some orbital phases (when the distance between the two stars is too close). As a result, depending on the stellar separation (that is largely varying in eccentric orbits) and on the orientation of the system (influencing the column of absorbing material along the line of sight), strong modulations are expected. This is in fact at the origin of strong biases in the identification of synchrotron radio emitters: a measurement at the wrong epoch may not reveal any synchrotron emission, despite the PACWB nature of the system.

Another consequence of an efficient DSA is the emission of $\upgamma$-rays with energies above several MeV. However, high-energy $\upgamma$-ray emission has only been reported from two PACWBs to date: the exceptional case of $\upeta$~Carinae \citep{tavani2009}, and in $\upgamma^2$~Velorum \citep[also known as WR~11;][]{martidevesa2020}. Recently, two more gamma-ray emitting candidates among PACWBs have been investigated through radio observations: HD~93129A \citep{benaglia2015} and Apep \citep{marcote2021} -- but, unfortunately, both remain undetected in gamma rays \citep{martidevesa2023}.

As it can be seen, the discovery of PACWBs encounter different challenges. These sources are highly variable, with a significant non-thermal emission, in may cases, limited to a fraction of the orbit. Additionally, many of these systems exhibit highly eccentric and long-period orbits (on the range of years or even tens of years). This parameter space remains poorly explored in terms of binary or higher multiplicity systems. Even optical surveys as {\em Gaia} would not reveal the binary of these systems in the near future.

\begin{figure}[!t]
	\centering
	\vspace{0pt}
	\includegraphics[width=\columnwidth]{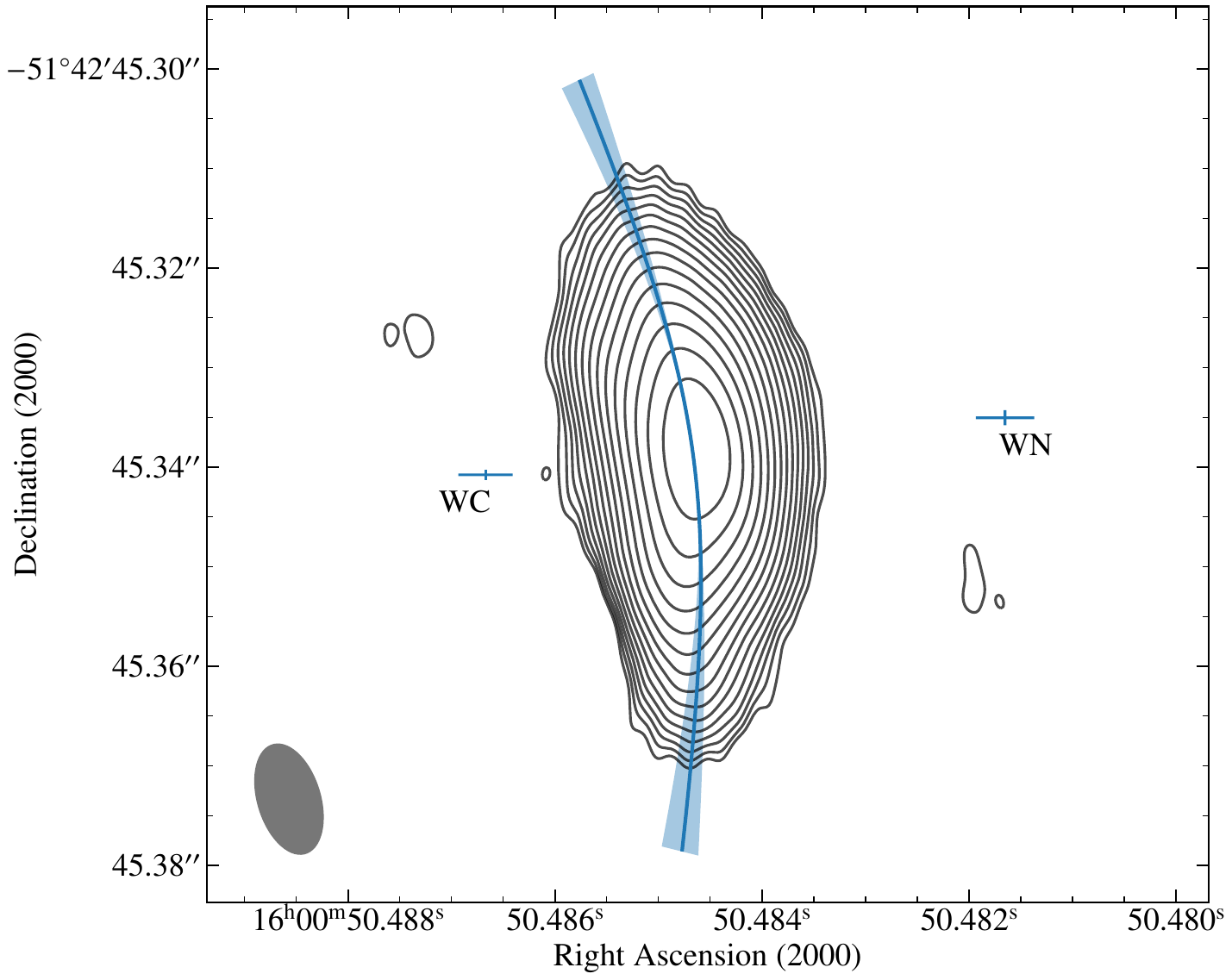}
	\caption{Radio image of the PACWB Apep as seen by the Australian Long Baseline Array \citep{marcote2021}. The contours represent the flux density above three times the rms noise level of $38~\mathrm{\upmu Jy~beam^{-1}}$, and increase by factors of $\sqrt{2}$. The curved solid blue line represents the contact discontinuity, that traces the shock produced between the stellar winds of the two stars in the binary system (a WC and a WN spectral type star). The positions of the stars, as derived from the contact discontinuity, are represented by the blue crosses (showing the $1\sigma$ uncertainty). The synthesized beam is $11.3 \times 5.6~\mathrm{mas^2}$, and is represented by the grey ellipse at the bottom left corner.}
	\label{fig:apep}
\end{figure}

However, radio interferometers may play an important role to unveil the unexplored population of PACWBs. Large field of view radio interferometers can monitor large fractions of the sky, specially those hosting complexes of massive stars, with a high cadence and a reduced observing time.
PACWBs would reveal themselves as non-thermal radio emitters when the conditions to produce DSA are adequate. In that moment, very high resolution radio observations via very long baseline interferometry (VLBI) would resolve the radio emitting region and unveil the bow-shape structure representative of a wind collision region. These images would ultimately confirm the PACWB nature of the system, as performed in the past \citep{benaglia2015,marcote2021}, see Fig.~\ref{fig:apep}.

Radio observations are thus key to unravel the mechanisms responsible for the non-thermal emission of PACWBs and gaining crucial insights into the dynamics of stellar collisions.

\subsection{Gamma Ray Binaries}

While we have already mentioned that there is a population of Galactic binaries emitting at HE or VHE, only a reduced number of them are energetic enough to consistently exhibit HE and VHE emission that dominates their non-thermal spectral energy distribution. This requires the presence of powerful mechanisms that accelerate particles up to relativistic energies, making these systems unique to study in detail particle acceleration given the short timescales involved and their relative proximity.

The systems verifying these properties are denominated gamma-ray binaries. To date we only know nine of such systems, and all of them are thought to comprise a massive star (or either O or B spectral type) and a young non-accreting pulsar or neutron star \citep{dubus2013,dubus2015}.

As the compact object interacts with the surrounding material and stellar wind of the massive star, a strong shock between the stellar wind and the relativistic pulsar wind is produced. In this medium is where high-energy particles are accelerated to relativistic velocities, boosting the production of gamma-rays. While the electrons escape the system and cool down, radio emission is displayed along the cometary tail that the shocked material originates as consequence of the orbital motion.  This non-thermal emission is highly variable, displaying periodic flares and quiescent phases in some cases, and provides insights into the complex dynamics of these systems.

Gamma-ray binaries offer a remarkable opportunity to study the extreme physics involved in the accretion processes, relativistic particle acceleration, and the interplay between the compact object and the stellar companion. Gamma-ray binaries represent one of the most extreme shocks observed in our Galaxy, and their dynamic and evolution can be monitored on day timescales. It is clear then that the study of these sources contributes to our understanding of high-energy astrophysics, the nature of particle acceleration, and the formation and evolution of compact binary systems in the universe.

\begin{figure}[!t]
	\centering
	\vspace{0pt}
	\includegraphics[width=\columnwidth]{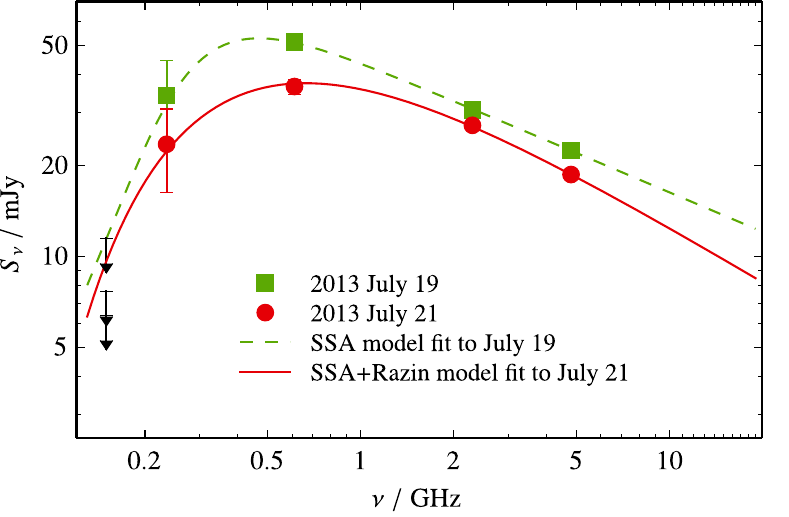}
	\caption{Spectra of the gamma-ray binary LS~5039 at two different epochs \citep{marcote2015}. In both cases, the power law from synchrotron emission is broken at low frequencies, exhibiting a turnover at a frequency of $\sim 0.5~\text{GHz}$. This turnover is produced due to absorption processes that dominate the spectrum. The best fits were obtained with a dominant component from synchrotron self-absorption. An extra component due to Razin effect (produced because of the surrounding stellar wind) was required in one of the two epochs to fully explain the observed spectrum.}
	\label{fig:ls5039}
\end{figure}

For the sources that have been explored in detail in the 0.1--5~GHz frequency range \citep{marcote2015,marcote2016}, it has been revealed how the absorption processes start to dominate at frequencies $\lesssim 1~\text{GHz}$, as shown in Fig.~\ref{fig:ls5039}. This allows us to directly constrain the turnover frequency of the non-thermal synchrotron radio emission, which in return allows the characterization of the physical properties of the emitting region, such as the average magnetic field, the electron density and temperature.

\section{Fast Radio Bursts}

In recent years, the discovery of Fast Radio Bursts (FRBs) has opened up a new frontier in transient astronomy.
FRBs are millisecond-duration radio flashes of coherent emission originated at cosmological distances with yet an unclear origin \citep[see][for a review]{petroff2022}.

Hundreds of FRBs have been discovered since the first detection by \citet{lorimer2007}, with estimated rates of $\sim 10^{3\text{--}4}~\mathrm{sky^{-1}~d^{-1}}$. The vast majority have been discovered by the CHIME/FRB telescope in Canada, which observes at 400--800~MHz. Among all these FRBs, only a small fraction seem to show multiple bursts --- the so-called {\em repeating} FRBs \citep{chen2022}.

Discovery of FRBs require large field of view searches, given the fact that the burst emission is not predictable. In addition, computationally expensive searches must be conducted through the high cadence raw data in order to detect a single burst. Most projects are motivated to detect a large number of FRBs, even with poor astrometry precision. Through statistical studies of the burst properties of FRBs, such as dispersion measures, polarization, and scattering profiles, researchers can provide unique probes for the the cosmic web, the nature of their progenitors, and even gain deeper insights into the fundamental physics of extreme astrophysical phenomena \citep{petroff2022}.

\begin{figure}[!t]
	\centering
	\vspace{0pt}
	\includegraphics[width=\columnwidth]{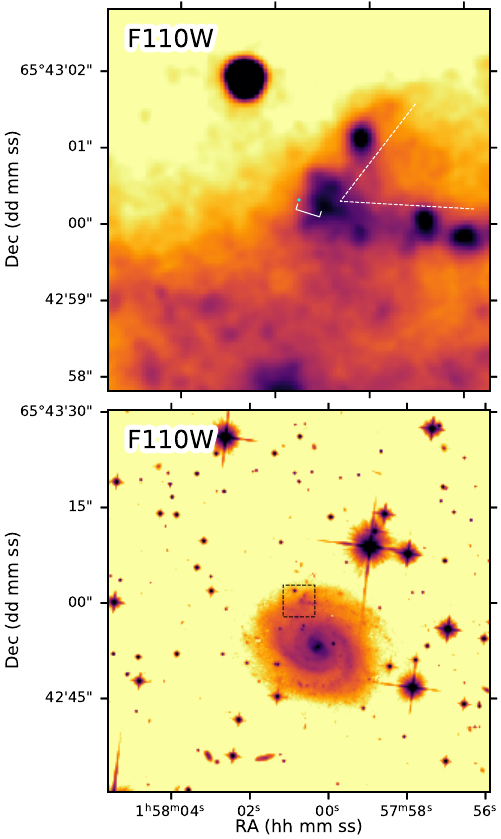}
	\caption{Precise localization of FRB~20180916B performed by the EVN and the {\em Hubble Space Telescope} \citep{marcote2020,tendulkar2021}. The milliarcsecond precision in the localization of the bursts (green dot in the top figure) allowed us to determine that the FRB was located right outside the edge of a prominent star-forming region of a spiral galaxy (shown in its full extension in the bottom panel. The top panel is a zoom in of the bottom one, covering the area represented by the dotted square.}
	\label{fig:frb}
\end{figure}

However, high angular resolution observations are mandatory to unveil the environments where FRBs are originated, and subsequently to understand the processes and physical conditions that lead to such luminous burst emission. Observations with the European VLBI Network (EVN) have proven to be a unique approach to achieve this goal. In the last years we have led the precise localizations of the first FRBs to milliarcsecond scales.
These localizations revealed a surprising fact about FRBs: they can be located in a variety of completely different environments and hosts galaxies. The first known repeating FRB, 20121102A, and also the first FRB to be localized with enough precision to be associated to its host galaxy, was found inside a star-forming region of a low-metallicity dwarf galaxy \citep{chatterjee2017,marcote2017,tendulkar2017,bassa2017}. And the following localized FRBs seemed to favor star-forming regions, even though the host galaxies were radically different \citep{marcote2020,tendulkar2021,nimmo2022}, see Fig.~\ref{fig:frb}.

However, we recently unveiled the localization of FRB~20200120E to the inside of a globular cluster associated to the M81 galaxy \citep{kirsten2022}.
This observed diversity seems to suggest that FRBs can be originated in a variety of environments and probably from different formation channels.

\section{The facilities in Argentina}

\subsection{The Multipurpose Interferometer Array (MIA)}

MIA is a currently-under-development project to build an interferometer that can operate between 100~MHz and 2~GHz with a maximum bandwidth of 1~GHz, and with up to 64 5-m antennas, in Argentina. With the proposed baselines of 50~m to 55~km, it will be able to operate with a maximum angular resolution of $\sim 1.5\arcsec$ and a field of view of about $4^\circ$ at 1.4~GHz.

Its design is particularly beneficial to investigate transient sources and non-thermal emission from both Galactic and extragalactic sources. As has been mentioned before, high-cadence large-field-of-view surveys can significantly contribute to increase the number of known Galactic binary systems. For populations that only consist of a handful or tens of sources, these new discoveries represent a step forward in the ability of performing statistical studies. Additionally, and at least in the case of PACWBs, the discovered population seems to be dominated by the exceptional (extreme) cases. To unveil the characteristics of the low-luminosity, more quiescent systems, observations on a larger scale are necessary.

MIA would represent a valuable facility for these kinds of observations. Targeting regions with high density of massive stars can reveal yet to know binary or higher multiplicity systems, specially those with elongated and long-period orbits.

In the case of FRBs, MIA could be proved to be a useful machine to discover a large number of bursts in the southern sky. We note that this hemisphere is still poorly sampled, as the current observatories searching for FRBs (e.g.\ ASKAP and MeerKAT) have a much limited field of view as compared to CHIME/FRB. The large field of view and low frequency range of MIA could potentially allow it to become the most prolific detector of FRBs in the Southern hemisphere. Additionally, if MIA would be able to search for bursts in the baseband (single-dish or phased-up) data, and still correlate the interferometric data after a FRB detection, it would imply that MIA could do both: detect a large number of FRBs and retrieve their positions in the sky with arcsecond or subarcsecond precision. This level of accuracy would be good enough to associate most of the discovered FRBs to their host galaxies.

\subsection{The 30-m IAR antennas}

IAR is also equipped with two 30-m radio antennas observing at 1.3 and 1.4~GHz with a bandwidth of 400~MHz \citep{gancio2020}. Successful observations in the last years, specially in the field of pulsar astronomy, have already been conducted with these antennas.

We have recently performed observations with one of the 30-m antennas on the gamma-ray binary PSR~B1259$-$63 during its periastron passage \citep[see][for more information about the source]{chernyakova2021}. These observations would reveal how the pulsed emission is affected by the decretion disk of the massive O9.5Ve-type companion star. The interaction between the pulsar when crossing such disk can be understood by monitoring the pulse signal during these moments. Changes in the dispersion measure (DM), Faraday rotation measure (RM), and the moment at which the pulses get completely suppressed due to free-free absorption, are direct indicators of the physical properties of the disk and the material where the pulsar is passing through.

\section{Discussion}

MIA is a promising array to study transient science. Depending on their final capabilities and their observing modes, it is expected to boost the transient science that is currently performed in the Southern hemisphere. A large number of Galactic binaries could be discovered with MIA if high cadence (likely with small dedicated observing time) observations along large fields are performed regularly. In particular, the location of MIA is much more convenient to explore the Galactic plane than northern instruments.

In the case of FRBs, MIA could demonstrate to be a perfect machine suitable to discover a large number of FRBs if the required components are in place: {\em 1)} being able to record data with high time and frequency resolution ($\sim \upmu\text{s}$ and $\lesssim \text{MHz}$, respectively), {\em 2)} a dedicated machine to perform single pulse searches on the data in a nearly real-time regime, and {\em 3)} being able to correlate the interferometric data once a burst is detected, to achieve a sub-arcsecond localization of the FRB.\\

The 30-m antennas have already proven their competitiveness in pulsar observations. However, and looking towards new paths for the near future, I would like to highlight the fact that they could, in addition, provide a remarkable contribution to the VLBI community. The current VLBI networks operating at gigahertz frequencies do not have any antenna located in South America. This leaves a significant gap on the long baselines, specially towards the North-South direction.

The current configuration of the 30-m antennas already makes them available to join such a network with a minimum effort.
Remarkable, the antennas already use a nearby H maser for timing, plus their current observing bandwidth is consistent with the typical bandwidth of VLBI observations (128--256~MHz, given that observations at 1.4 or 1.6~GHz are conducted at  1--2~Gbps, using 2-bit sampling, with the EVN).
Only internal conversions to record the data into an appropriate format (e.g.\ VDIF; \citealt{whitney2010}), likely with minimal changes in the backend system, would be necessary.

The inclusion of one of the 30-m antennas in the EVN would provide the second long baseline towards the South direction, together with the Hartebeesthoek antenna in South Africa. In addition, it would represent the longest baseline in the array (12\,360~km, with Urumqi, in China; both overlapping for a very short period of time for a source located at the celestial equator).
With respect to the core of the EVN antennas, located in Europe, the 30-m antenna at IAR would provide a reasonable $\sim 6\text{-h}$ overlap (again, for a equatorial source).

This path should thus be considered and the community would benefit for such inclusion, specially for all Galactic sources that are located near the Galactic plane.

\section{Conclusions}

The study of transient sources at gigahertz radio frequencies with interferometers provides an extraordinary opportunity to explore the intricate astrophysical processes at play in our dynamic Universe. Gamma-ray binaries, colliding wind binaries, and the rapidly-evolving field of Fast Radio Bursts present captivating realms of inquiry. The project to place a competitive interferometer like MIA, and the recent developments of the 30-m antennas in Argentina, would harness the studies of transient sources in an hemisphere that typically remains poorly explored as compared to the Northern one. Such facilities would allow us to unlock the mysteries that remain about these sources and pave the way for groundbreaking discoveries that will enhance our understanding of the cosmos.

\end{document}